\documentclass{elsart}
\usepackage{times}
\usepackage{epsfig}
\usepackage{marvosym}
\usepackage{amssymb}
\usepackage{url}
\begin{document}
\bibliographystyle{roman}

\journal{Nuclear Instruments and Methods A}

\def\hb{\hfill\break}
\def\MeV{\rm MeV}
\def\GeV{\rm GeV}
\def\TeV{\rm TeV}

\def\m{\rm m}
\def\cm{\rm cm}
\def\mm{\rm mm}
\def\lam{$\lambda_{\rm int}$}
\def\rad{$X_0$}
 
\def\NIM{Nucl. Instr. and Meth.~}
\def\ieee {{IEEE Trans. Nucl. Sci.~}}
\def\prl{Phys. Rev. Lett.~}

\def\etal{{\it et al.}}
\def\eg{{\it e.g.,~}}
\def\ie{{\it i.e.,~}}
\def\cf{{\it cf.~}}
\def\etc{{\it etc.}}
\def\vs{{\it vs.~}}

\hyphenation{ca-lo-ri-me-ter}
\hyphenation{ca-lo-ri-me-ters}
\hyphenation{Brems-strah-lung}

\begin{frontmatter}
\title{On the limits of the hadronic energy resolution of calorimeters*}

\author{Sehwook Lee$^a$, Michele Livan$^b$ and Richard Wigmans$^{c,}$\thanksref{Corres}}

\address{$^a$ Department of Physics, Kyungpook National University, Daegu, Korea\\
$^b$ Dipartimento di Fisica, Universit\`a di Pavia and INFN Sezione di Pavia, Italy\\
$^c$ Department of Physics and Astronomy, Texas Tech University, Lubbock (TX), USA}
\thanks[Corres]{Corresponding author.
              Email wigmans@ttu.edu, fax (+1) 806 742-1182.}
              
\vskip -7mm
\begin{abstract}
In particle physics experiments, the quality of calorimetric particle detection is typically considerably worse for hadrons than
for electromagnetic showers. In this paper, we investigate the root causes of this
problem and evaluate two different methods that have been exploited to remedy this situation: compensation
and dual readout. It turns out that the latter approach is more promising, as evidenced by experimental results.

\vskip -5mm
\end{abstract}
\end{frontmatter}
{\sl * This paper is dedicated to the memory of our long-time friend and collaborator Guido Ciapetti}
\newpage
\section{Introduction}
\vskip -5mm
In the past half-century, calorimeters have become very important components of the detector system at almost every experiment in high-energy particle physics. This is especially true for $4\pi$ experiments at high-energy particle colliders, such as LEP and the Large Hadron Collider at CERN, the Tevatron at Fermilab and RHIC at Brookhaven. Experiments at proposed future colliders such as the FCC (CERN), CEPC (China) and ILC (Japan) will be designed around a powerful central calorimeter system.

A calorimeter is a detector in which the particles to be detected are completely absorbed. The detector provides a signal that is a measure for the energy deposited in the absorption process. In {\sl homogeneous} calorimeters, the entire detector volume may contribute to the signals. In {\sl sampling} calorimeters, the functions of particle absorption and signal generation are exercised by different materials, called the {\sl passive} and the {\sl active} medium, respectively. Almost all calorimeters operating in the mentioned experiments are of the latter type. The passive medium is usually a high-density
material, such as iron, copper, lead or uranium. The active medium generates the light or charge
that forms the basis for the signals from such a calorimeter.

Among the reasons for the increased emphasis on calorimetric particle detection in modern experiments, we mention
\begin{itemize}
\item The fact that calorimeters can provide important information on the particle collisions, in particular information on the {\sl energy flow}
in the events (transverse energy, missing energy, jet production, \etc) 
\item Calorimeters can provide this information {\sl very fast}, almost instantaneously. In modern experiments, \eg at the LHC, it has become possible to decide whether an event is worth retaining for offline inspection on a time scale of the order of $10^{-8}$ seconds. Since the LHC experiments have to handle event rates at the level of 10$^9$ each second, this triggering possibility is a crucial property in these experiments.
\item Calorimeter data can be very helpful for {\sl particle identification}.   
\item Important aspects of the calorimeter performance, such as the energy and position resolutions, tend to {\sl improve with energy}.
\end{itemize}

Calorimetric detection of $\gamma$'s and electrons has a long tradition, which goes back to the early days of nuclear spectroscopy,
when scintillating crystals such as NaI(Tl) were the detectors of choice. In high-energy physics, detection of electromagnetic showers 
is nowadays routinely performed with a resolution at the 1\% level, both in homogeneous \cite{bgo} and sampling \cite{na48} calorimeters. 

The success of experiments at a future high-energy $e^+e^-$ Collider will also depend critically on the quality of the hadron calorimetry. Unfortunately, the performance of hadron calorimeters leaves much to be desired.

In this paper, we describe first the reasons for the generally poor performance of calorimeters intended to detect hadrons and jets (Section 2). In Sections 3 and 4, two methods that have been developed as a remedy for these problems are presented, and the performance improvement achieved with these methods is compared
in Section 5. Conclusions are given in Section 6.

\section{The problems of hadron calorimetry}
\vskip -5mm
The development of hadronic cascades in dense matter differs in essential ways from that of electromagnetic ones, with important consequences for calorimetry.
Hadronic showers consist of two distinctly different components:
\begin{enumerate}
\item An {\sl electromagnetic} component; $\pi^0$s and $\eta$s generated in the absorption process
decay into $\gamma$'s which develop em showers.
\item A {\sl non-electromagnetic} component, which combines essentially everything else that
takes place in the absorption process. 
\end{enumerate}
For the purpose of calorimetry, the main difference between these components is that
some fraction of the energy contained in the non-em component does {\sl not} contribute to the signals. This {\em invisible energy}, which mainly consists of the binding energy of nucleons released in the numerous nuclear reactions, may represent up to 40\% of the total non-em energy, with large event-to-event fluctuations.
\begin{figure}[b!]
\epsfxsize=10cm
\centerline{\epsffile{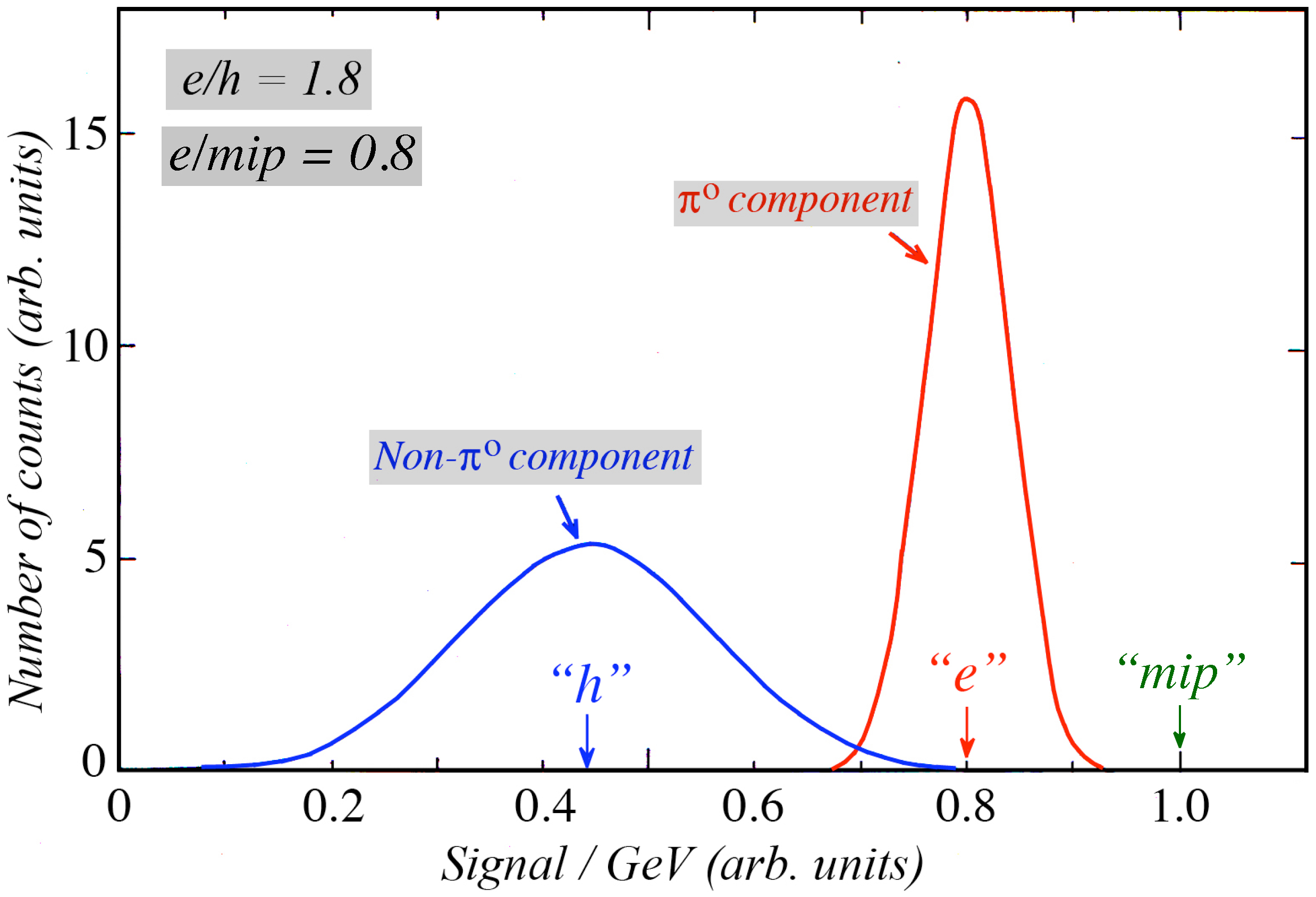}}
\caption{\small
Illustration of the meaning of the $e/h$ and $e/mip$ values of a calorimeter. Shown are distributions of the signal per unit deposited energy for the electromagnetic and non-em components of hadron showers. These distributions are normalized to the response for minimum ionizing particles ($``mip"$). The average values of the em and non-em distributions are the em response ($``e"$) and non-em response ($``h"$) , respectively.}  
\label{ehprinciple}
\end{figure}

Let us define the calorimeter {\em response} as the conversion efficiency from deposited energy to generated signal, and normalize it to electrons. The responses of a given calorimeter to the em and non-em hadronic shower components, $e$ and $h$, are usually not the same, as a result of invisible energy and a variety of other effects. We will call the distribution of the signal per unit deposited energy around the mean value ($e$ or $h$) the {\sl response function}.

Figure \ref{ehprinciple} illustrates the different aspects of the calorimeter response schematically. The em response is larger than the non-em one, and the non-em response function is broader than the em one, because of event-to-event fluctuations in the invisible energy fraction. Both $e$ and $h$ are smaller than the calorimeter response for minimum ionizing particles, because of inefficiencies in the shower sampling process \cite{Liv17}. The calorimeter is characterized by the $e/h$ and $e/mip$ ratios, which in this example have values of 1.8 and 0.8, respectively.
Calorimeters for which $e/h \ne 1$ are called {\sl non-compensating}.
\begin{figure}[htbp]
\epsfxsize=14cm
\centerline{\epsffile{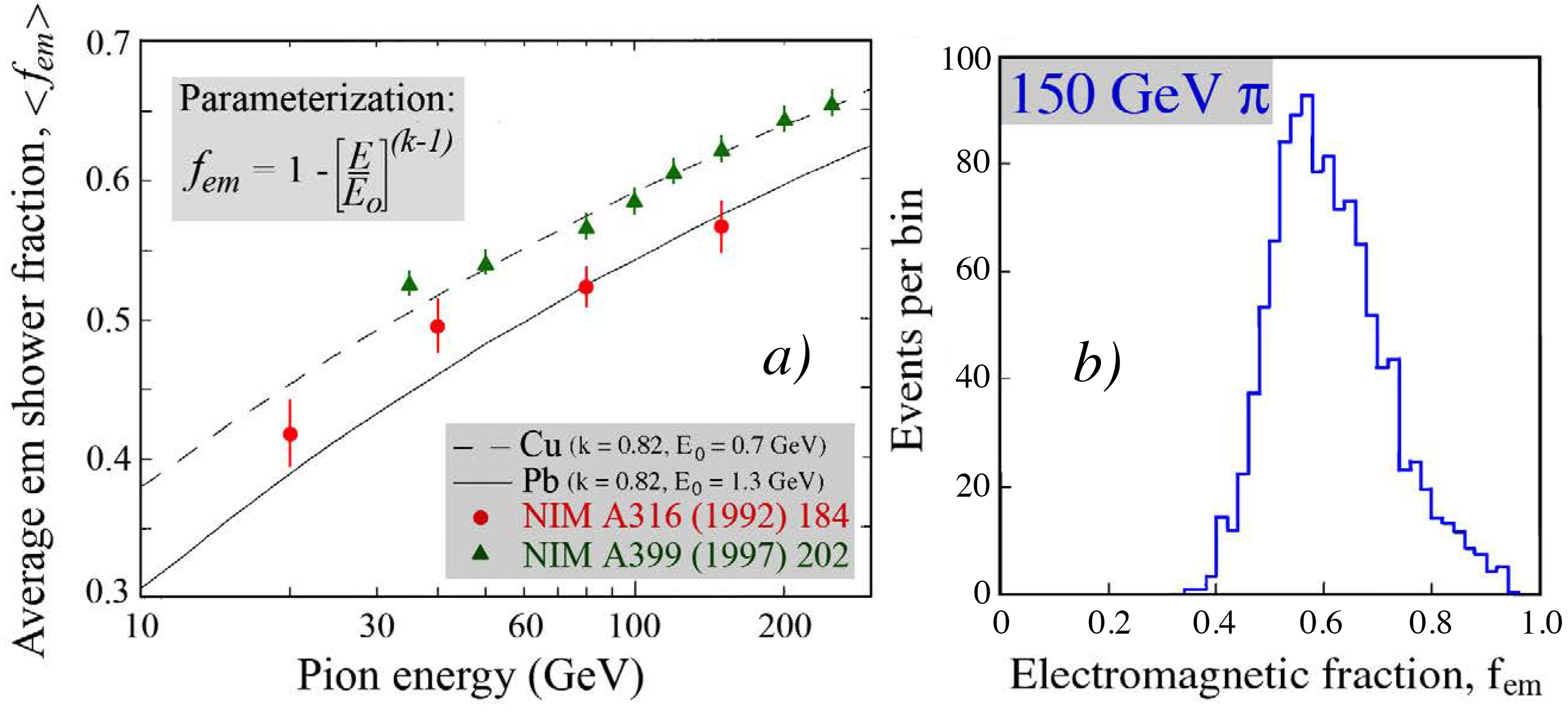}}
\caption{\small
Properties of the electromagnetic fraction of hadron showers. Shown are the measured values of the average value of that fraction as a function of energy, for showers developing in lead or copper ($a$) and the distribution of $f_{\rm em}$ values measured for 150 GeV $\pi^-$ showers developing in lead ($b$). The curves in diagram $a$ represent Equation \ref{femE}. Experimental data from \cite{Aco92b,Akc97}.} 
\label{femprops}
\end{figure}

\begin{figure}[b!]
\epsfxsize=14cm
\centerline{\epsffile{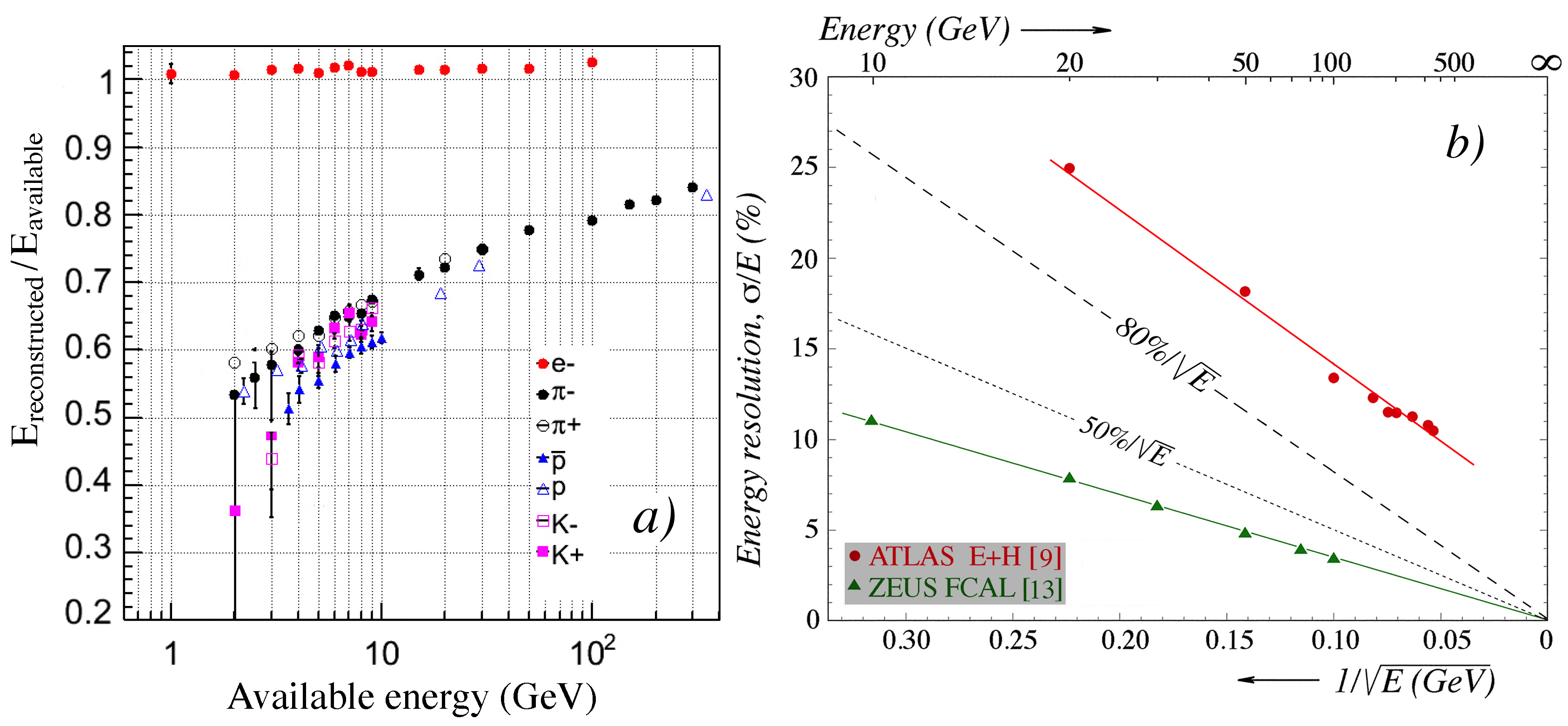}}
\caption{\small
Experimental consequences of non-compensation for the hadronic calorimeter performance. The non-linearity reported by CMS ($a$) and the energy resolution reported by ATLAS ($b$). For comparison, the hadronic energy resolution reported for the compensating ZEUS calorimeter is shown as well. See text for details.}  
\label{detperformance}
\end{figure}

The properties of the em shower component have important consequences for the hadronic {\em energy resolution}, signal {\em linearity} and {\em response function}.
The average fraction of the total shower energy contained in the em component, $\langle f_{\rm em} \rangle$, was measured to increase
with energy following a power law \cite{Aco92b,Akc97}, confirming an induction argument made to that effect \cite{Gab94}:
\begin{equation}
\langle f_{\rm em} \rangle ~=~1 - \biggl[ \biggl({E\over E_0}\biggr)^{k-1}\biggr]
\label{femE}
\end{equation}
where $E_0$ is a material-dependent constant related to the average multiplicity in hadronic interactions (varying from 0.7 GeV to 1.3 GeV for $\pi$-induced reactions on Cu and Pb, respectively), and $k \sim 0.82$ (Figure \ref{femprops}a). For proton-induced reactions, $\langle f_{\rm em} \rangle$ is typically considerably smaller, as a result
of baryon number conservation in the shower development \cite{Akc98}.
A direct consequence of the energy dependence of $\langle f_{\rm em} \rangle$ is that calorimeters for which $e/h \ne 1$ are by definition {\sl non-linear} for hadron detection, since the response to hadrons is given by 
$ \langle f_{\rm em} \rangle + \bigl[1 -  \langle f_{\rm em} \rangle\bigr] h/e$. %
This is confirmed by many sets of experimental data, for example the ones reported for CMS \cite{CMS07} shown in Figure \ref{detperformance}a.

Event-to-event fluctuations in $f_{\rm em}$ are large and non-Poissonian \cite{Aco92b}, as illustrated in Figure \ref{femprops}b. If $e/h \ne 1$, these fluctuations 
tend to dominate the hadronic energy resolution and their asymmetric characteristics are reflected in the 
response function \cite{Liv17}.
It is often assumed that the effect of non-compensation on the energy resolution is energy independent (``constant term''). This is incorrect, since it implies that the effect is insignificant at low energies, \eg 10 GeV, which is by no means the case. The measured effects of {\em fluctuations} in $f_{\rm em}$ can be described by a term that is very similar to the one used for its energy dependence
(\ref{femE}). 
This term should be added in quadrature to the $E^{-1/2}$ scaling term which accounts for all Poissonian fluctuations:
\begin{equation}
{\sigma\over E}~=~{a_1\over \sqrt{E}} \oplus a_2 \biggl[ \biggl({E\over E_0}\biggr)^{l-1}\biggr]
\label{lognoncomp}
\end{equation}
where the parameter $a_2 = |1 - h/e|$ is determined by the degree of non-compensation \cite{deg07}, and $l \sim 0.72$.
It turns out that in the energy range covered by the current generation of test beams, \ie up to 400 GeV, 
Equation \ref{lognoncomp} leads to results that are very similar to those from an expression of the type
\begin{equation}
{\sigma\over E}~=~{c_1 \over \sqrt{E}} + c_2
\label{linres}
\end{equation}
\ie a {\em linear sum} of a stochastic term and a constant term. Many sets of experimental hadronic energy resolution data exhibit exactly this characteristic, for example the results reported for ATLAS \cite{Aba10} shown in Figure \ref{detperformance}b. In this figure, the energy resolution is plotted 
on a scale linear in $E^{-1/2}$, inverted to increase from right to left\footnote{This convention is used for all energy resolution plots in this paper.}. Scaling with $E^{-1/2}$ is thus represented by a straight line through the bottom right corner in this plot. The experimental ATLAS data are located on a line that runs parallel to such a line, indicating that the stochastic term ($c_1$) is $\approx 80\%$ and the constant term ($c_2$) is $\approx 5\%$ in this case.

The root cause of the poor performance of hadron calorimeters is thus the invisible energy. Because some fraction of the energy carried by the hadrons 
and released in the absorption process does not contribute to the signal, the response to the non-em shower component is typically smaller than that to the em shower component. And the characteristic features of the energy sharing between these two components lead to hadronic signal non-linearity, a poor energy resolution and a non-Gaussian response function.

To mitigate these effects, one  thus needs a measurable quantity that is correlated to the invisible energy. The stronger that correlation, the 
better the hadronic calorimeter performance may become. In the next two sections, two such measurable quantities are discussed: the kinetic energy released by neutrons in the absorption process (section 3) and the total non-em energy (section 4).

\section{Compensation}
\vskip -5mm
The first successful attempt to mitigate the effects described in the previous section involved a calorimeter that used depleted uranium as absorber material. The underlying idea was that the fission energy released in the absorption process would compensate for the invisible energy losses. By boosting the non-em calorimeter response in this way, the $e/h$ ratio would increase and, as a matter of good fortune, reach the (ideal) value of 1.0. This is the reason why calorimeters with $e/h = 1.0$ have become known as {\sl compensating} calorimeters. 

Indeed, it turned out that the mentioned effects  of non-compensation on energy resolution, linearity and line shape, as well as the associated calibration problems \cite{CMS07} are absent in compensating calorimeters. 
However, it also turned out that fission had nothing to do with this, and that the use of uranium was neither necessary nor sufficient
for reaching the compensation condition. The crucial element was rather the active material of the sampling calorimeter, which 
had to be very efficient in detecting the numerous neutrons produced in the shower development process. Hydrogenous active material
may meet that condition, since in a sampling calorimeter with high-$Z$ passive material, MeV type neutrons lose most of their kinetic energy in elastic neutron-proton scattering, whereas the charged particles are sampled according to $dE/dx$.
 
Compensation can thus be achieved in sampling calorimeters with high-$Z$ absorber material and hydrogenous active material. It requires a very specific sampling fraction, so that the response to shower neutrons is boosted by the precise factor needed to equalize $e$ and $h$. For example, in Pb/scintillating-plastic structures, this sampling fraction is $\sim 2\%$ for showers \cite{Bern87,Aco91c,Suz99}. This small
sampling fraction sets a lower limit on the contribution of sampling fluctuations to the energy resolution, while the need to efficiently detect MeV-type neutrons requires signal integration over a relatively large volume and at least 30 ns. Yet, the experiment that holds the current world record in hadronic energy resolution (ZEUS, $\sigma/E \sim 35\%/\sqrt{E}$) used a calorimeter of this type \cite{Beh90}). The experimental energy resolution data reported for this calorimeter are shown in Figure \ref{detperformance}b. Especially at high energies, this resolution is much better than that of the calorimeters currently operating in the LHC experiments.

In compensating calorimeters, the total kinetic energy of the neutrons produced in the hadronic shower development thus represents the measurable quantity correlated to the invisible energy. The relative magnitude of the signal provided by these neutrons can be tuned to achieve equality of the electromagnetic and non-electromagnetic calorimeter responses ($e./h = 1.0$), by means of the sampling fraction. This mechanism works because the calorimeter response to charged shower particles is much more sensitive to a change in the sampling fraction than the response to neutrons. 

\section{Dual-readout calorimetry}
\vskip -5mm
The dual-readout approach aims to achieve the advantages of compensation without the
disadvantages mentioned in the previous section:
\begin{itemize}
\item The need for high-$Z$ absorber material, and the associated small $e/mip$ value, which causes non-linearities at low energy
and deteriorates the jet performance \cite{zeusjet},
\item A small sampling fraction, which limits the em energy resolution,
\item The need to detect MeV-type neutrons efficiently, which implies integrating the signals over relatively large detector volumes and long times.
\end{itemize}

The purpose of the dual-readout technique is to measure the em shower fraction ($f_{\rm em}$) {\sl event by event}. If successful, this would make it possible to diminish/eliminate the effects of fluctuations in $f_{\rm em}$ on the hadronic calorimeter performance.
This was in itself not a new idea. Starting around 1980, attempts have been made to disentangle the energy deposit profiles of hadronic showers with the goal to identify the em components, which are typically characterized by a high localized energy deposit \cite{Abr81}.
Such methods are indeed rather successful for isolated high-energy hadron showers, but fail at low energy, and in particular in cases where a number of particles develop showers in the same vicinity, as is typically the case for jets.  

The dual-readout method exploits the fact that
the energy carried by the non-em shower components is mostly deposited by non-relativistic shower particles (protons), and therefore does not contribute to the signals of a 
\v{C}erenkov calorimeter. By measuring simultaneously $dE/dx$ and the \v{C}erenkov light generated in the shower absorption process, one can determine $f_{\rm em}$ event by event and thus eliminate
(the effects of) its fluctuations. The correct hadron energy can be determined from a combination of both signals.

This principle was first demonstrated by the DREAM Collaboration \cite{Akc05a}, with a Cu/fiber calorimeter. Scintillating fibers measured $dE/dx$, quartz fibers the \v{C}eren-kov light.
The response ratio of these two signals is related to $f_{\rm em}$ as
\begin{equation}
{C\over S} ~=~ {{f_{\rm em} +  (1 - f_{\rm em}) (h/e)_C}\over {f_{\rm em} +  (1 - f_{\rm em}) (h/e)_S}}
\label{eq2}
\end{equation}
where $(h/e)_C = 0.21$ and $(h/e)_S = 0.77$ are the values of the $h/e$ ratios for the \v{C}erenkov and scintillator structures in the DREAM calorimeter, respectively. In general, the em fraction is thus given by:
\begin{equation}
f_{\rm em} ~=~ {{(C/S) (h/e)_S - (h/e)_C}\over {[1 - (h/e)_C] - (C/S) [1 - (h/e)_S]}}
\label{eq2a}
\end{equation}

The hadron energy ($E$) can be derived directly from the two signals $S$ and $C$ \cite{deg07}:
\begin{equation}
E~=~ {{S - \chi C}\over {1 - \chi}}
\label{eq5}
\end{equation}
in which $\chi$ is constant, independent of energy and of the particle type, determined solely by the $e/h$ values of the scintillation and \v{C}erenkov calorimeter structures:
\begin{equation}
\chi~=~ {{1 - (h/e)_S}\over {1 - (h/e)_C}}
\label{eq6}
\end{equation}
\begin{figure}[b!]
\epsfxsize=14cm
\centerline{\epsffile{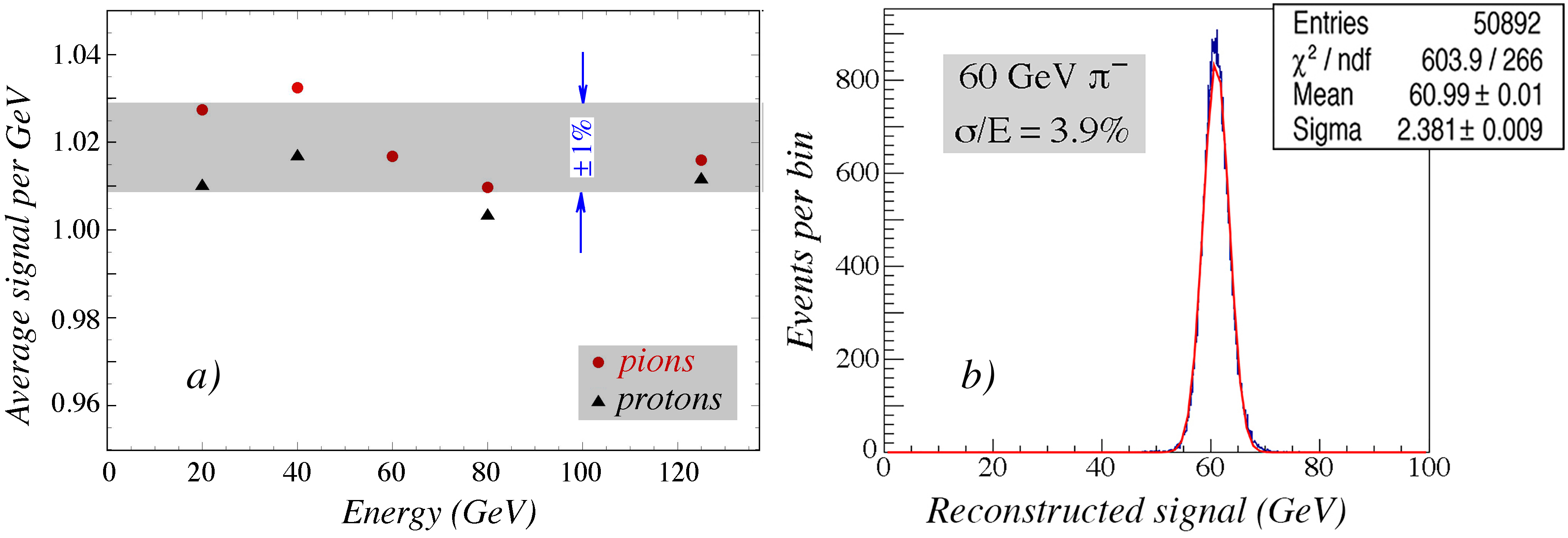}}
\caption{\small
Results obtained with the RD52 lead-fiber dual-readout calorimeter \cite{slee17}. Shown are the average signal per unit deposited energy as a function of energy, separately for pions and protons ($a$), and the measured signal distribution for 60 GeV $\pi^-$ ($b$).} 
\label{rd52}
\end{figure}

Some of the merits of this method are illustrated in Figure \ref{rd52}, which shows that the dual-readout calorimeter is very linear and produces the same response for pions and protons (Figure \ref{rd52}a), that the response function is well described by a Gaussian and, most importantly,
that the hadronic energy was correctly reproduced in this way (Figure \ref{rd52}b). This was true both for single pions as well as for multiparticle events \cite{slee17}. In the Appendix, details are given about the particular way in which results such as those shown in Figure \ref{rd52} were obtained.

In dual-readout calorimeters, the total non-em energy, which can be derived from the measured total energy (Equation \ref{eq5}) and the em shower fraction (Equation \ref{eq2}), thus represents the measurable quantity correlated to the invisible energy. The limitations that apply for compensation do not apply in this case. Any absorber material may be used, as a matter of fact the dual-readout method may even be applied for {\sl homogeneous} calorimeters, such as BGO crystals \cite{Akc09c}. The sampling fraction is not restricted and neutron detection is not a crucial ingredient for this method.
Therefore, one is considerably less constrained when designing a calorimeter system of this type than in case of a system based on compensation.

\section{Dual-readout \vs compensation}
\vskip -5mm
Compensating calorimeters and dual-readout calorimeters both try to eliminate/\break
mitigate the effects of fluctuations in the invisible energy on the signal distributions by means of a measurable variable that is correlated to the invisible energy. As mentioned in the previous sections, the variables used for this purpose are different in compensating and dual-readout calorimeters. However, with both methods a very significant improvement of the hadronic calorimeter performance is obtained, compared to the standard non-compensating calorimeters used in the current generation of particle physics experiments:
the hadronic response is constant (\ie the calorimeter is linear for hadron signals), the hadronic response function Gaussian, the hadronic energy resolution much better and, most importantly, a calibration with electrons also provides the correct energy 
for hadronic showers.

\begin{figure}[htbp]
\epsfysize=8.5cm
\centerline{\epsffile{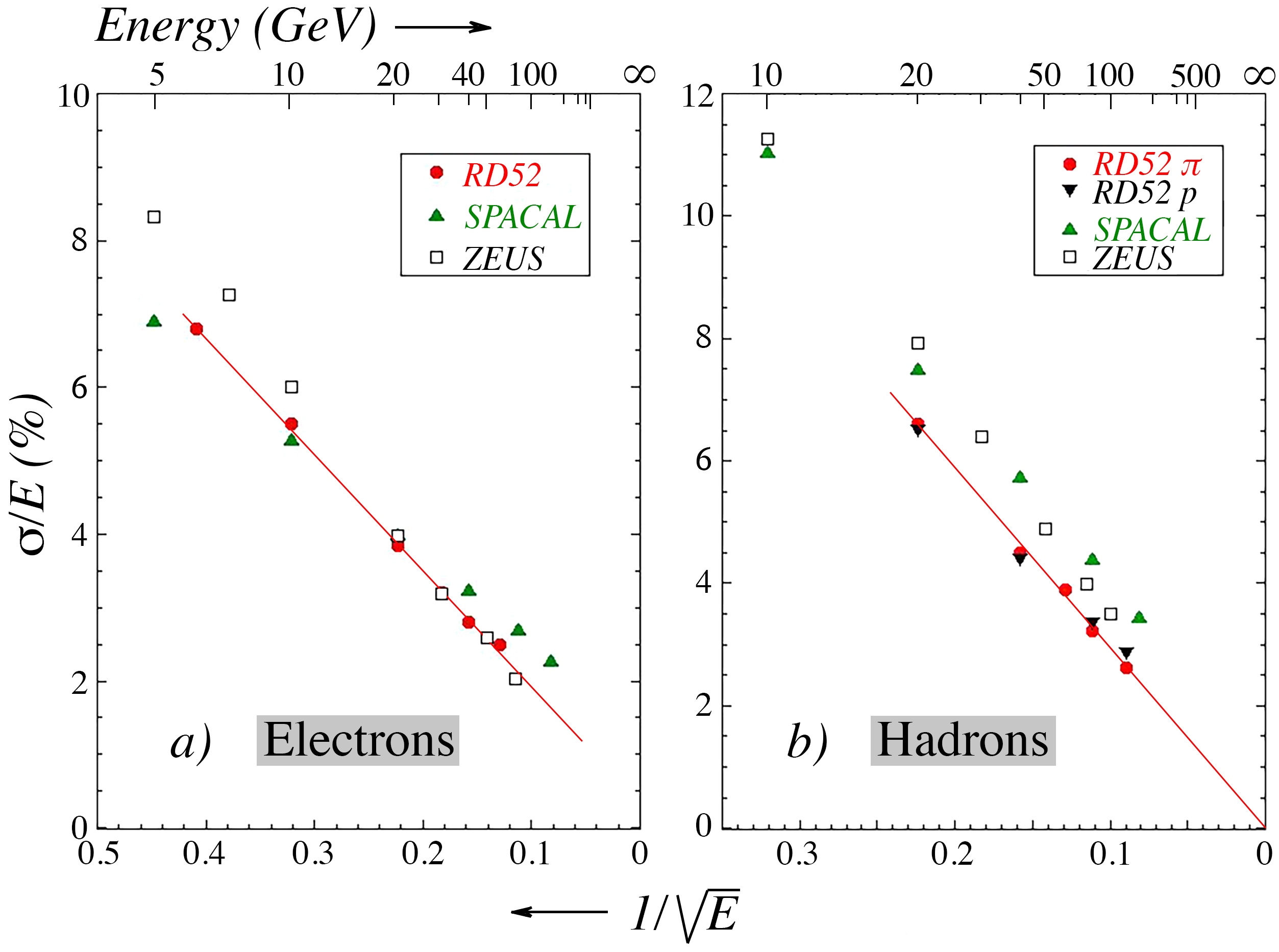}}
\caption{\small
Energy resolutions reported for the detection of electrons($a$) and hadrons ($b$) by RD52 \cite{slee17,Akc14}, SPACAL \cite{Aco91c} and ZEUS \cite{Beh90}.}
\label{Erescomp} 
\end{figure}

In Figure \ref{Erescomp}, we compare the energy resolutions obtained with the best compensating calorimeters, ZEUS \cite{Beh90} and SPACAL \cite{Aco91c}, and the results obtained with the RD52 dual-readout fiber calorimeter. Figure \ref{Erescomp}b shows that the hadronic RD52 values are actually better than the ones reported by ZEUS and SPACAL, while Figure \ref{Erescomp}a shows that the RD52 em energy resolution is certainly not worse.

In making this comparison, it should be kept in mind that
\begin{enumerate}
\item The em energy resolutions shown for RD52 were obtained with the calorimeter oriented at a much smaller angle with the beam line 
($\theta,\phi = 1^\circ ,1.5^\circ$) than the ones for SPACAL ($\theta,\phi = 2^\circ ,3^\circ$) \cite{Akc14}. It has been shown that
the em energy resolution is extremely sensitive to the angle between the beam particles and the fiber axis when this angle is very small \cite{Car16}.
\item The instrumented volume of the RD52 calorimeter (including the leakage counters) was less than 2 tons, while both SPACAL and ZEUS
obtained the reported results with detectors that were sufficiently large ($> 20$ tons) to contain the showers at the 99+\% level. The hadronic resolutions shown for RD52 are dominated by fluctuations in lateral shower leakage, and a larger instrument of this type is 
thus very likely to further improve the results.
\end{enumerate}
%

The comparison of the hadron results (Figure \ref{Erescomp}b) seems to indicate that the dual-readout approach offers better opportunities to achieve superior hadronic performance than compensation. 
Apparently, in hadronic shower development the correlation with the total nuclear binding energy loss is thus stronger for the 
total non-em energy (derived from the em shower fraction) than for the total kinetic neutron energy. Intuitively, this is not a surprise, since the total non-em energy consists of other components than just neutrons, and the total kinetic energy of the neutrons is 
not an exact measure for the {\sl number} of neutrons (which is the parameter expected to be correlated to the binding energy loss).

In order to investigate the validity of this interpretation of the experimental results, we performed Monte Carlo simulations of shower development in a block of matter that was sufficiently large to make the effects of longitudinal and lateral shower leakage insignificantly small\footnote{Leakage through the front face (albedo) cannot be avoided, but this is an effect at the level of a fraction of 1\% for multi-GeV hadron absorption.}. Large blocks of copper or lead were used for this purpose.

The simulations were carried out with the GEANT4 Monte Carlo package \cite{geant}.
Events were generated with GEANT4.10.3 patch-02, which was released in July 2017. 
For applications of calorimetry in high energy physics, GEANT4 recommends to use the FTFP\_BERT physics list which contains the Fritiof model~\cite{Fritiof}, coupled to the Bertini-style cascade model~\cite{Bertini} and all standard electromagnetic processes. 
This is the default physics list used in simulations for the CMS and ATLAS experiments at CERN's Large Hadron Collider~\cite{g4_pl}.

Pions of different energies were absorbed in these structures. For each event, the following information was extracted: 
\begin{enumerate}
\item The {\sl em shower fraction}, $f_{\rm em}$.  This was determined by summing the energy carried by $\pi^0$s and $\gamma$'s produced in the shower development and comparing it with the total energy ($E_\pi$) of the showering pion. The total non-em energy was then determined as 
$E_{\rm non-em} = E_\pi (1 - f_{\rm em})$.
\item The {\sl total kinetic neutron energy}, $E_{\rm kin}(n)$. For this purpose, all neutrons produced in nuclear reactions in the shower development were 
taken into account. A special effort was made to avoid potential double-counting issues, for example when a spallation neutron initiated a nuclear reaction in which other neutrons were produced. 
\item The {\sl total nuclear binding energy loss}, $\Delta B$. This was determined from the final-state nuclei produced in the shower development.
In order to limit the complications of this approach, we used target material consisting of only one (the most abundant) isotope. Simulations on copper or lead were thus carried out using $^{63}$Cu or $^{208}$Pb as target material.
\end{enumerate}

Simulations were carried out for pions of 10, 20, 50 and 100 GeV. For each run, 10,000 events were generated. These (time consuming) simulations yielded a lot of information. In situations where the results could be compared to experimental data, such as those shown in Figure \ref{femprops}, the agreement was good.
A detailed comparison with all available experimental data is foreseen to be the topic of a forthcoming paper. For our purpose, we were primarily interested in the correlation between
the nuclear binding energy loss, which is the main culprit for poor hadronic calorimeter performance, and the variables devised to mitigate the effects of that, \ie the total kinetic neutron energy or $f_{\rm em}$.

\begin{figure}[htb]
\epsfysize=5cm
\centerline{\epsffile{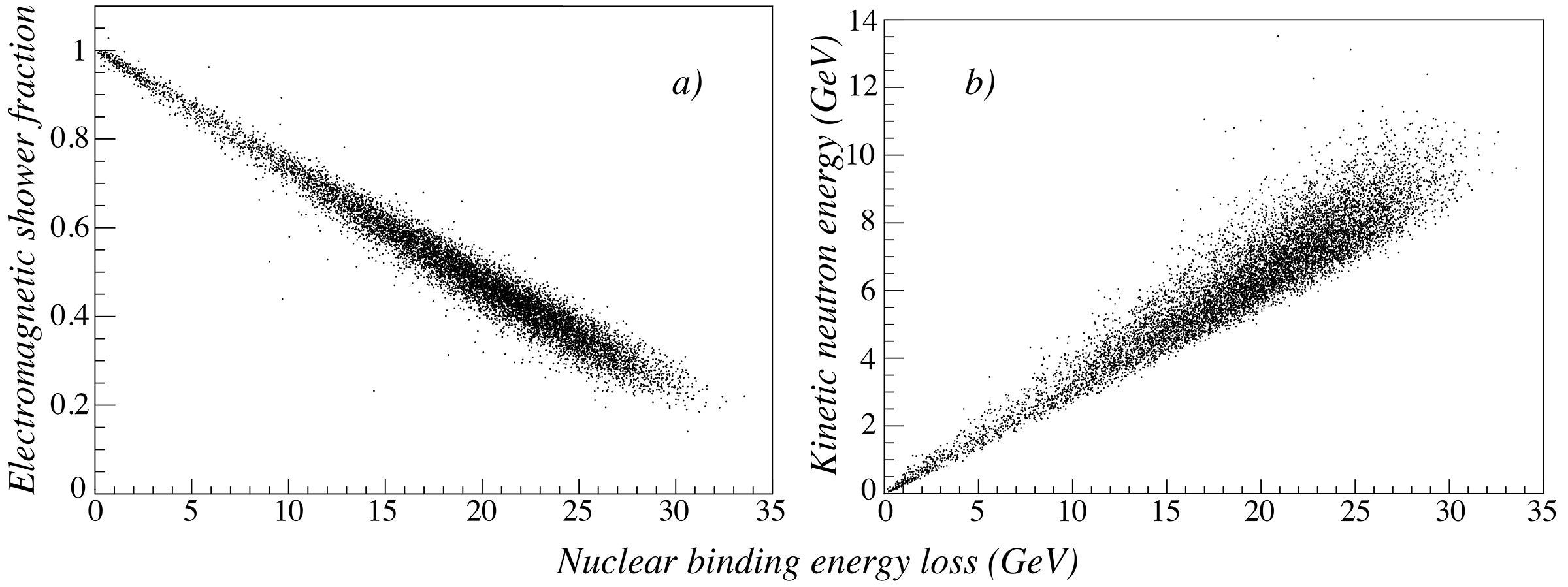}}
\caption{\small
The em fraction of the deposited energy ($a$) and the total kinetic energy carried by neutrons ($b$) are plotted versus the total nuclear binding energy loss when 100 GeV $\pi^-$ are absorbed in a massive block of lead. Results from GEANT Monte Carlo simulations. }
\label{sehwook1}
\end{figure}

\begin{figure}[htb]
\epsfysize=5.2cm
\centerline{\epsffile{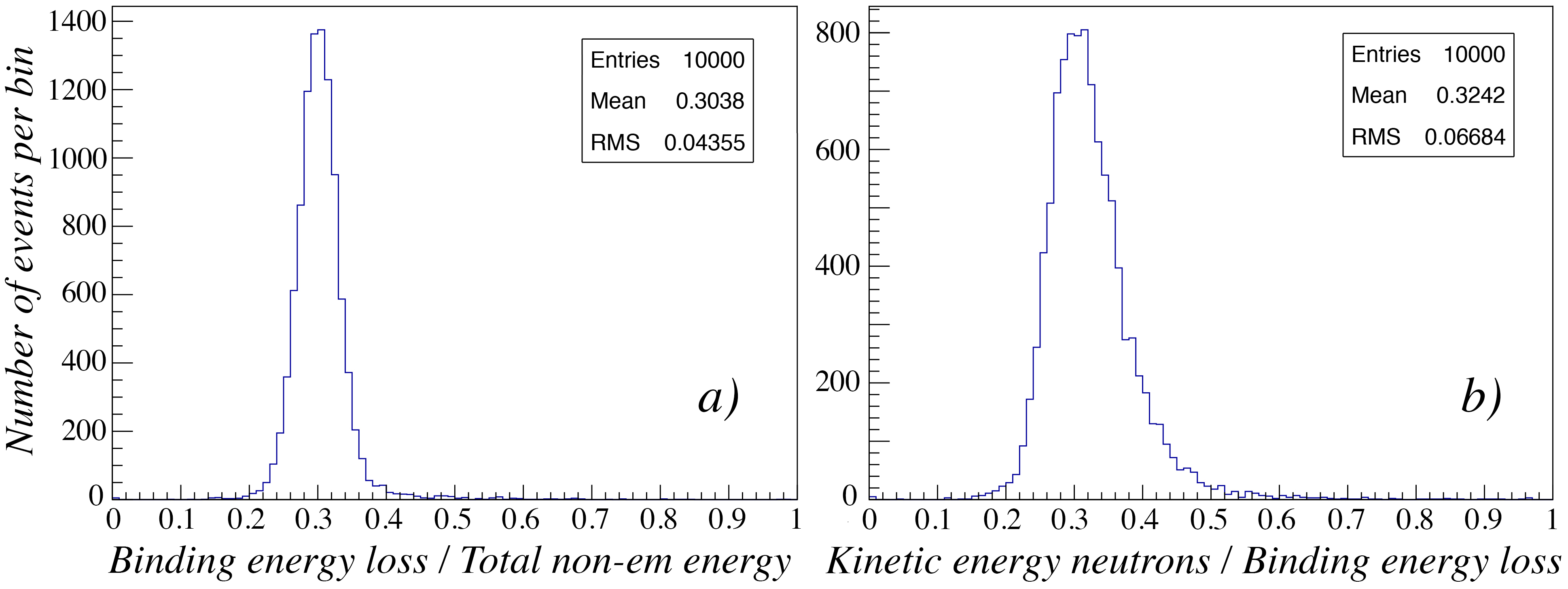}}
\caption{\small
Distributions of the ratio of the non-em energy and the nuclear binding energy loss ($a$) and the ratio of the total kinetic energy carried by neutrons and the nuclear binding energy($b$) for hadron showers generated by 50 GeV $\pi^-$ in a massive block of copper. Results from GEANT Monte Carlo simulations. }
\label{sehwook2}
\end{figure}

\begin{figure}[htb]
\epsfysize=7cm
\centerline{\epsffile{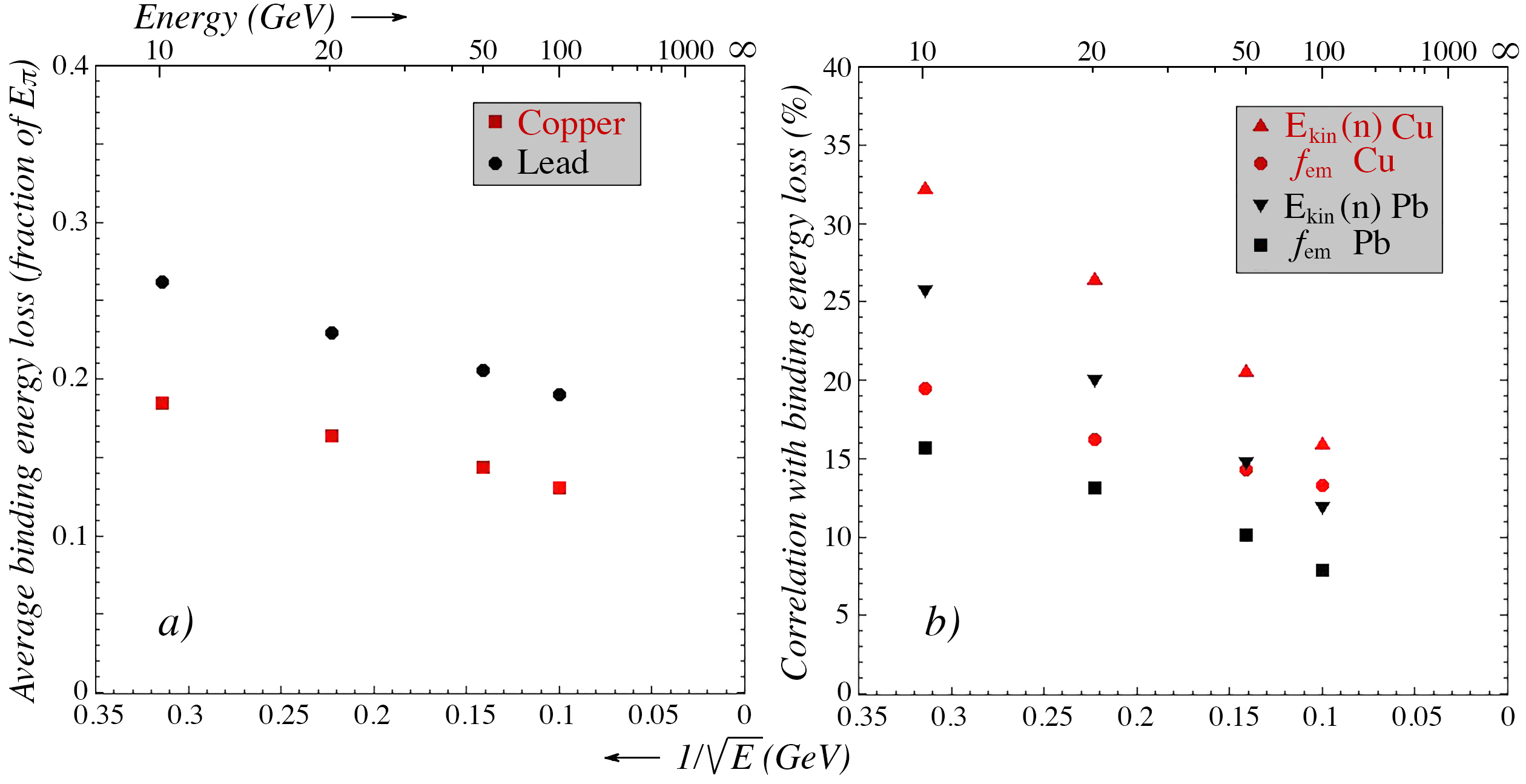}}
\caption{\small
The average energy fraction spent on nuclear binding energy losses for pion showers developing in a block of copper or lead, as a function of the pion energy ($a$).
The correlation between the em shower fraction and this nuclear binding energy loss and the correlation between the total kinetic neutron energy and this nuclear binding energy loss as a function of energy for pion showers developing in a massive block of copper or lead ($b$). Results from GEANT Monte Carlo simulations.}
\label{sehwook3}
\end{figure}

Figure \ref{sehwook1} shows some results of these simulations for 100 GeV $\pi^-$ absorbed in lead. 
The em energy fraction ($a$) and the total kinetic neutron energy ($b$) are plotted versus the total nuclear binding energy loss, in a scatter plot where each event is represented by one dot.  
Both the em shower fraction and the total kinetic neutron energy are clearly correlated with the nuclear binding energy loss.
To examine the degree of correlation, event-by-event ratios were determined. For this purpose, the em shower fraction was replaced by the non-em energy,
which has the same degree of correlation with $\Delta B$ as $f_{\rm em}$, but has the same energy dependence as $\Delta B$. The total non-em energy is measured event by event in dual-readout 
calorimeters, since it represents a fraction (1 - $f_{\rm em}$) of the particle energy.
Histograms of these ratios are shown in Figure \ref{sehwook2} for 50 GeV $\pi^-$ showers in copper, and the {\sl rms/mean} value of these distributions is plotted in Figure \ref{sehwook3}b as a function of the pion energy for both copper and lead absorber. 

These figures confirm that the correlation between the total non-em energy and the nuclear binding energy loss is better than the correlation between
the total kinetic neutron energy and the nuclear binding energy loss. This is true both for copper and for lead. For lead, the correlations are somewhat better than for copper. For the total kinetic neutron energy, this is to be expected, since in lead most of the nucleons released in the nuclear reactions are actually neutrons. 
However, the correlation with the total non-em energy is also better in lead. This is a consequence of the fact that in high-$Z$ absorber material
a larger fraction of the available shower energy is dissipated in the form of nuclear reactions, rather than the production of pions and other mesons \cite{Gab94}. The GEANT4 Monte Carlo simulations confirmed this aspect as well, as illustrated by Figure \ref{sehwook3}a, which shows the average fraction of the energy of the incoming pion that is used to break up atomic nuclei. This fraction is considerably larger for lead than for copper.

The results shown in Figure \ref{sehwook3} can be used to estimate the effects of the correlations discussed above on the energy resolution for hadron 
calorimeters that are based on dual-readout or compensation. This is done as follows. For example at 20 GeV, the average binding energy loss in copper 
amounts to 16.4\% of the energy of the incoming pion. The correlation between $\Delta B$ and $E_{\rm non-em}$ is 16.2\%, which means that the pion energy can be determined with a precision of $0.164 \times 0.162 = 2.7\%$ in that case. The correlation between $\Delta B$ and $E_{\rm kin}(n)$ is 26.5\%, which gives an energy resolution for the 20 GeV pions of 4.3\%.
\begin{figure}[htb]
\epsfysize=8.5cm
\centerline{\epsffile{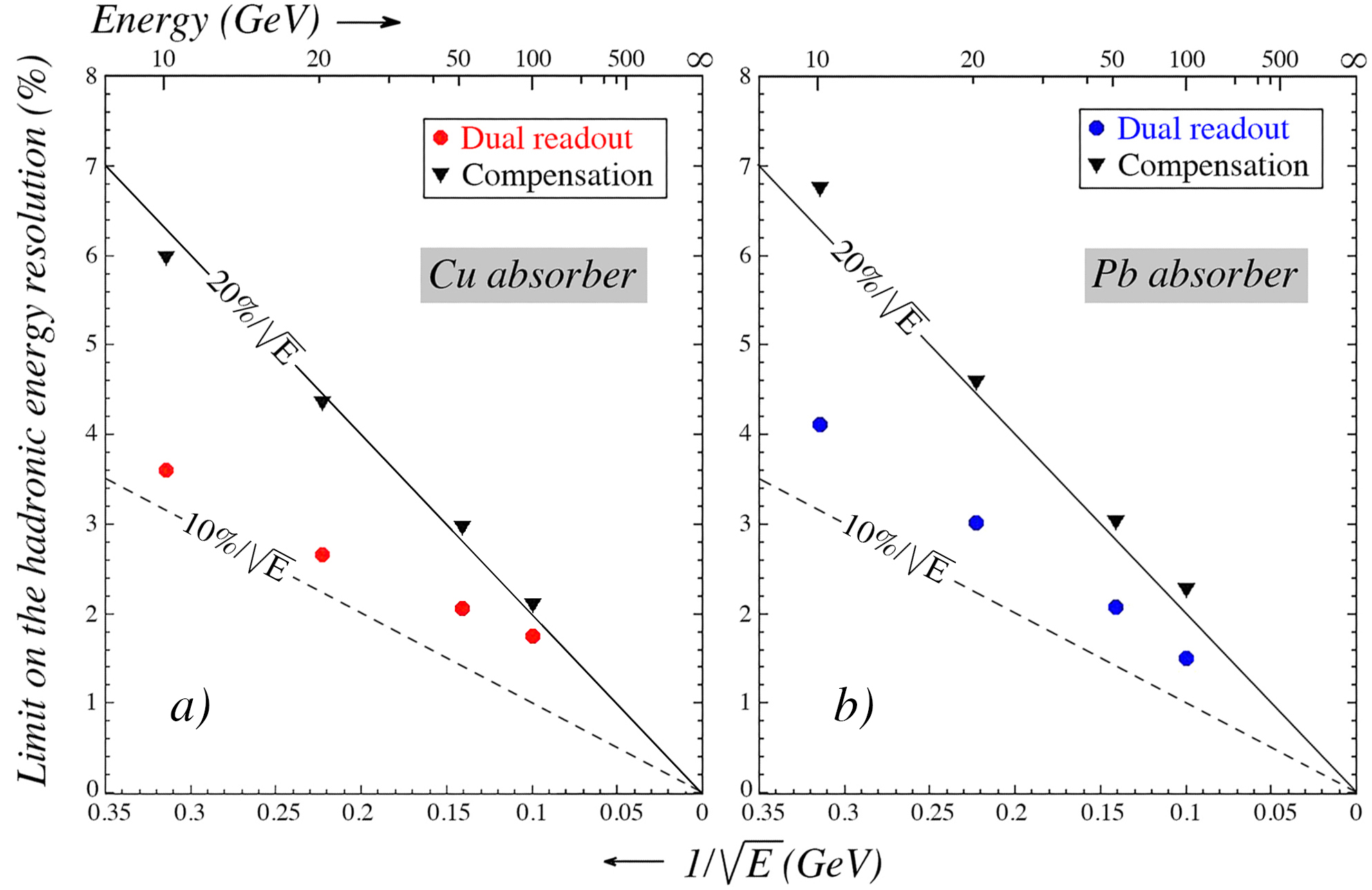}}
\caption{\small
The limit on the hadronic energy resolution derived from the correlation between nuclear binding energy losses and the parameters measured in dual-readout or compensating calorimeters, as a function of the particle energy. The straight lines represent resolutions of $20\%/\sqrt{E}$ and $10\%/\sqrt{E}$, respectively, and are intended for reference purposes.
Results from GEANT Monte Carlo simulations of pion showers developing in a massive block of copper ($a$) or lead ($b$).}
\label{sehwook4}
\end{figure}

These results are summarized in Figure \ref{sehwook4}, for pions in copper ($a$) and lead ($b$). Interestingly, the differences between these two absorber materials shown in Figure \ref{sehwook3} approximately cancel each other, and the results are very similar for the two absorber materials. 
These resolutions should be considered Monte Carlo predictions for the {\sl ultimate hadronic energy resolution} that can be achieved with calorimeters
using either dual-readout or compensation as the method to mitigate the effects of (fluctuations in) invisible energy. 
For reference purposes, we have drawn lines corresponding to resolutions of $10\%/\sqrt{E}$ and $20\%/\sqrt{E}$ in this figure. 
The limits for compensating calorimeters scale remarkably well with $E^{-1/2}$, while the limits for dual-readout calorimeters exhibit a small deviation of this scaling behavior. However, the latter limits are considerably better than for compensating calorimeters, at all energies considered here.

\vskip 3mm
\begin{figure}[htbp]
\epsfysize=8cm
\centerline{\epsffile{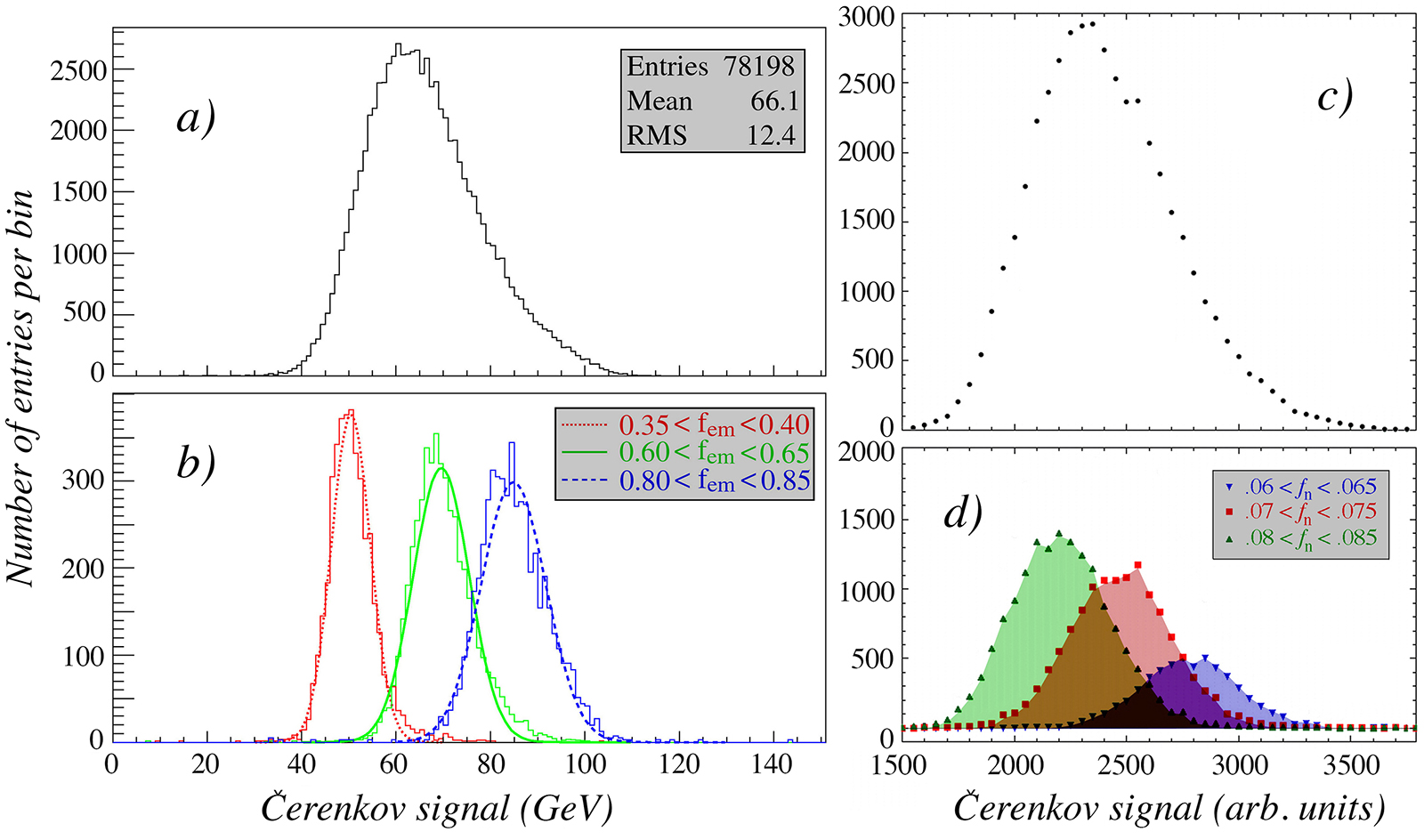}}
\caption{\small
Distribution of the total \v{C}erenkov signal for 100 GeV $\pi^-$ ($a$) and the distributions for three subsets of events selected on the basis of the electromagnetic shower fraction ($b$). Data from \cite{Akc05a}.
Distribution of the total \v{C}erenkov signal for 200 GeV multiparticle events ($c$) and the distributions for three subsets of events selected on the basis of the fractional contribution of neutrons to the scintillator signal ($d$). Data from \cite{Akc09a}.}
\label{femnextract}
\end{figure}

Experimental data obtained by the RD52 Collaboration also support our conclusion that the correlation exploited in dual-readout calorimeters provides a more accurate measurement of the invisible energy. Figure \ref{femnextract}a,b shows that the (\v{C}erenkov) signal from the DREAM
fiber calorimeter is actually a superposition of many rather narrow, Gaussian signal distributions. Each sample in Figure \ref{femnextract}b contains events with (approximately) the same $f_{\rm em}$ value, \ie with the same total non-em energy. 
The dual-readout method combines all these different subsamples and centers them around the correct energy value. The result
is a relatively narrow, Gaussian signal distribution with the same central value as for electrons of the same energy.

Figure \ref{femnextract}d shows that the DREAM (\v{C}erenkov) signal is also a superposition of Gaussian signal distributions of a different type. In this  case, each sample consists of events with (approximately) the same total kinetic neutron energy. The dual-readout method may combine all these different subsamples in the same way as described above.
In doing so, the role of the total non-em energy is taken over by the total kinetic neutron energy, and the method becomes thus very similar to the one used in compensating calorimeters.

A comparison between Figures \ref{femnextract}b and \ref{femnextract}d shows that the signal distributions from the event samples are clearly wider when the total kinetic neutron energy is chosen to dissect the overall signal. This is consistent with our assessment that dual-readout is a more effective way to reduce the effects of fluctuations in invisible energy on the hadronic energy resolution.

Apart from that, dual-readout offers also several other crucial advantages:
\begin{itemize}
\item Its use is not limited to high-$Z$ absorber materials.
\item The sampling fraction can be chosen as desired.
\item The performance does not depend on detecting the neutrons produced in the absorption process. Therefore, there 
is no need to integrate the calorimeter signals over a large detector volume.
\item The signal integration time can be limited for the same reason. 
\end{itemize}
\begin{figure}[htbp]
\epsfxsize=14cm
\centerline{\epsffile{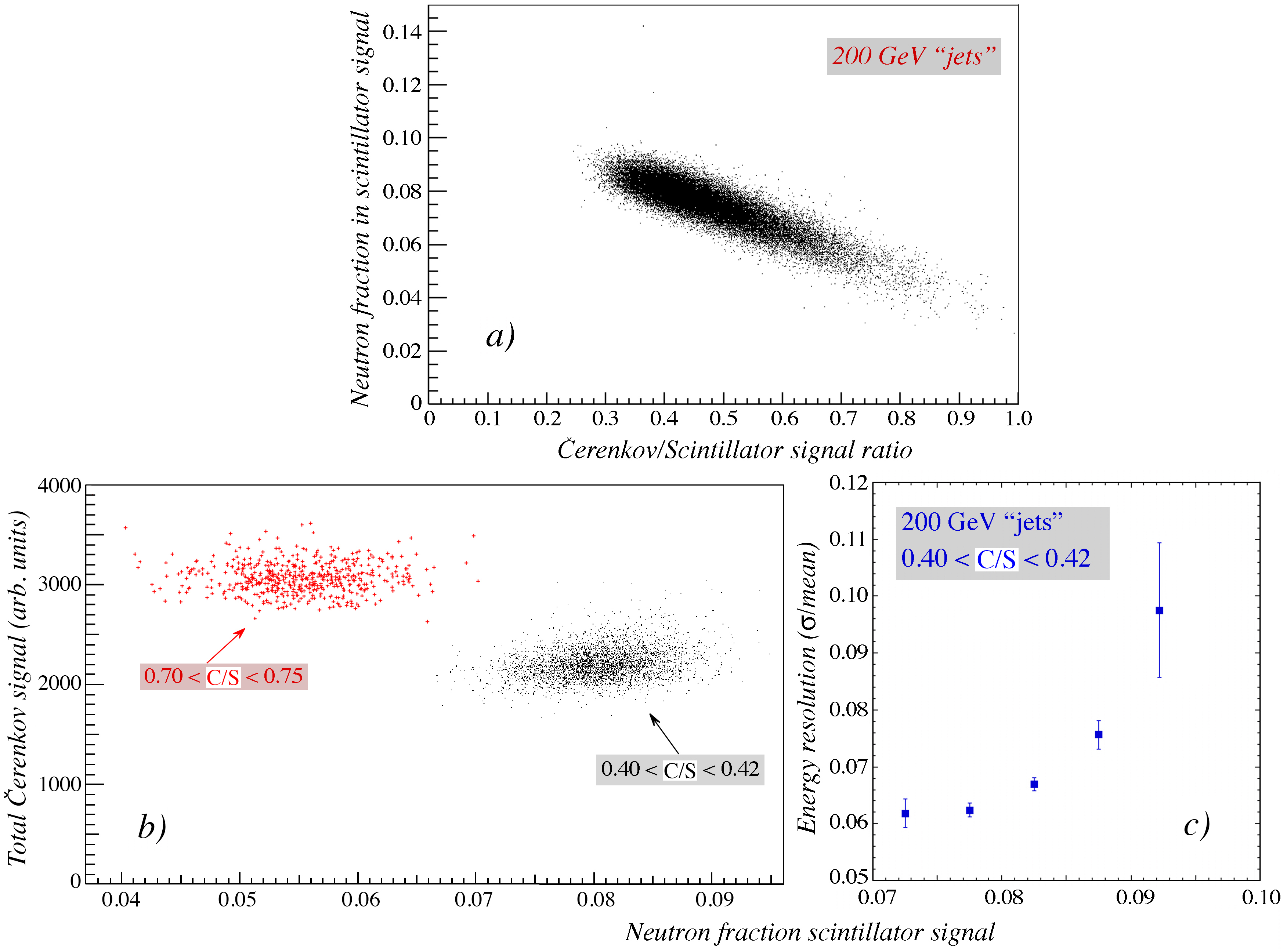}}
\caption{\small
Results obtained with the DREAM copper-fiber dual-readout calorimeter, in which the time structure of the signals was measured \cite{Akc09a}.
A scatter plot in which the measured contribution of neutrons to the signals ($f_n$) is plotted versus the measured ratio ($C/S$) of the \v{C}erenkov and scintillation signals ($a$). A scatter plot in which the measured \v{C}erenkov signal is plotted versus $f_n$, for two different bins in the $C/S$ distribution ($b$). The energy resolution for different  $f_n$ values, for events with (approximately) the same $f_{\rm em}$ value ($c$). }  
\label{femfncor}
\end{figure}

This is not to say that there is no advantage in detecting the neutrons produced in the shower development. In fact, this may further improve the hadronic calorimeter resolution, since $f_{\rm em}$ and $f_n$ are correlated with the nuclear binding energy losses in different ways, and thus may offer complementary benefits. Figure \ref{femfncor}a shows that a decrease in the \v{C}erenkov/scintillation signal ratio (from which $f_{\rm em}$ can be derived) corresponds to an increase of the neutron component ($f_n$) of the scintillation signal. However, as shown in Figure \ref{femfncor}b, this correlation is not perfect. In this scatter plot, the $f_n$ values are plotted for two narrow bins in the distribution of the \v{C}erenkov/scintillation signal ratio. In both cases, the $f_n$ values cover a much larger range than the $\pm 2\%$ range of the $f_{\rm em}$ values. Figure \ref{femfncor}c shows that the energy resolution depends
rather strongly on the chosen $f_n$ value, for a given value of $f_{\rm em}$. RD52 has shown that the complementary information provided by measurements of $f_{\rm em}$ and $f_n$ leads to a further improved hadronic energy resolution \cite{Akc09a}. However, even without explicitly determining $f_n$, which involves measuring the time structure of each and every signal, the hadronic energy resolution that can be obtained with the dual-readout method is already superior to what has been achieved by the best compensating calorimeters.

\section{Conclusions}
\vskip -5mm

The hadronic performance of the calorimeter systems currently used in experiments at high-energy particle colliders is dominated by fluctuations in the energy fraction used to break up atomic nuclei in the shower development. Two different methods have been proposed and tested to mitigate these effects: compensation and dual-readout. Both methods have been demonstrated to be very effective in improving the hadronic calorimeter performance.
Calorimeters based on these methods have achieved hadronic signal linearity, Gaussian response functions, very good hadronic energy resolutions and a correct reconstruction of the hadronic energy in instruments calibrated with electrons. We have investigated and compared the principles on which both methods are based and concluded that dual-readout calorimetry provides somewhat superior performance, combined with fewer practical restrictions, than compensation.

\section*{Acknowledgements}
\vskip -5mm

This study was carried out with financial support of the United States
Department of Energy, under contract DE-FG02-12ER41783, of Italy's Istituto Nazionale di Fisica Nucleare and Ministero dell'Istruzione, dell' Universit\`a e della Ricerca, and of
the Basic Science Research Program of the National Research Foundation of Korea (NRF), funded by the Ministry of Science, ICT \& Future Planning under contract 2015R1C1A1A02036477. 

\section*{Appendix}
\vskip -5mm

\begin{figure}[hbt]
\epsfysize=12.5cm
\centerline{\epsffile{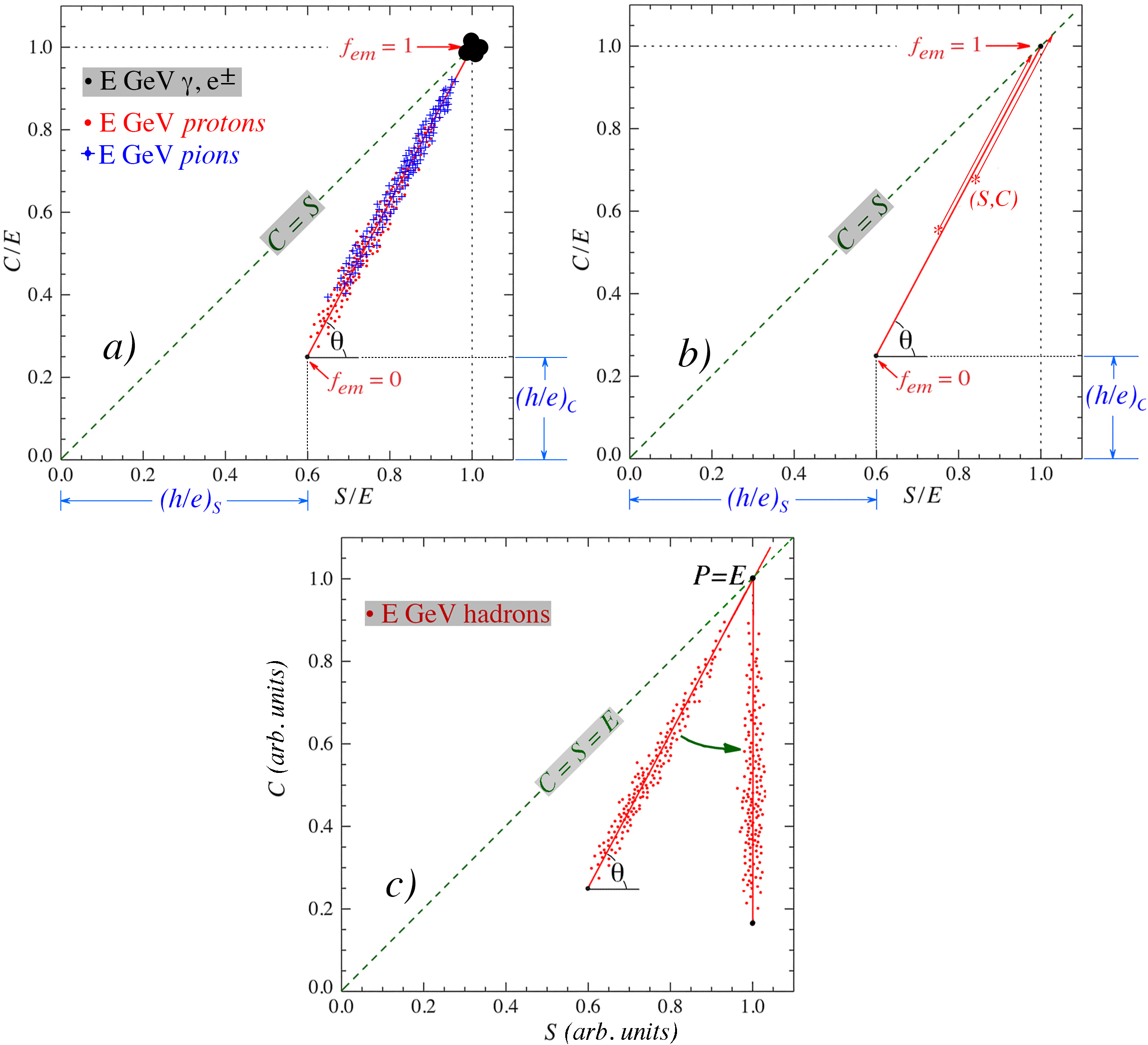}}
\caption{\small
The $S-C$ diagram of the signals from a dual-readout calorimeter. The hadron events are clustered around the straight (red) line that links the points for which  $f_{\rm em} = 0$ and $f_{\rm em} = 1$. Protons and pions have different $f_{\rm em}$ distributions and are thus clustered differently around this line ($a$). The effect of the dual-readout formula (\ref{eq5}) is shown in diagram $b$.
The effect of a rotation of the red line and the associated distribution of data points (\ref{rot}) is shown in diagram $c$. }
\label{drot}
\end{figure}

The performance results of the RD52 dual-readout calorimeter shown in Figure \ref{rd52} were obtained with a method described in this Appendix.
As pointed out by D. Groom in the Review of Particle Physics \cite{PDG16}, the ($S,C$) data points from a dual-readout calorimeter are clustered around a straight line in the scatter plot (Figure \ref{drot}a). This line links the point for which $f_{\rm em} = 0$ with the point for which $f_{\rm em} = 1$.
The latter point is located on the diagonal ($C = S$), where also the data points for em showers are located. 
The angle $\theta$ is only determined by the $e/h$ values of the \v{C}erenkov and scintillation calorimeter structures and is {\sl independent} of energy and of the particle type. It is also the same for single hadrons and jets. Its value is related to the $\chi$ parameter in Equation \ref{eq6}, as: $\chi = \cot{\theta}$. This equation represents a transformation in which each data point is moved up along the red line until it intersects with the diagonal, as illustrated in Figure \ref{drot}b. The data points obtained in this way are thus clustered around the same energy value as the data points for electrons of the same energy. Projecting these data points on the horizontal ($S$) and vertical ($C$) axes leads to Gaussian signal distributions centered around the correct particle energy.  This is true for both pions and protons, even though the $f_{\rm em}$ distributions for these particles may be quite different
(Figure \ref{drot}a). 

The projections of the data points obtained with Equation \ref{eq5} on the $S$ and $C$ axes are identical, since these data points are now located on the diagonal. However, the energy resolution of the calorimeter is better than the width of these distributions, which is dominated by fluctuations in the \v{C}erenkov light yield. The limiting factor for the hadronic energy resolution of this calorimeter is rather determined by the fluctuations of the data points in the scatter plot around the (red) line that links the points for which $f_{\rm em} = 0$ and $f_{\rm em} = 1$. 

The effects of these fluctuations on the signal distributions become clear when the measured distribution of data points is rotated around 
point $P$, which is the intersection of this line and the diagonal. 
If the calorimeter is calibrated with electrons, then $P$ is the point around which all electron showers that carry the same energy as the hadrons are clustered. In other words, the coordinates of $P$ reveal the energy of the hadron beam.
The rotation of the measured distribution of the hadronic data points around $P$, over an angle $90^\circ - \theta$, corresponds to a coordinate transformation of the type
\begin{equation}
\left(\matrix{S^\prime \cr C^\prime\cr}\right)~=~
\left(\matrix{\sin{\theta}&-\cos{\theta}\cr \cos{\theta}&\sin{\theta}\cr}\right)
\left(\matrix{S\cr C\cr}\right)
\label{rot}
\end{equation}
The distribution shown in Figure \ref{rd52}b is the projection of this rotated distribution on the horizontal ($S$) axis. Not surprisingly, the projection of the data points on the vertical ($C$) axis becomes much broader after this rotation.

In this way, the information contained in the two signals provided by the calorimeter has been used to obtain the most precise information about the energy of the beam particles. We want to stress the fact that the energy of the beam particles has {\sl not} been used in this procedure. That energy was obtained from a straight-line fit through the locus of experimental data points. The intersection of this line and the diagonal of the
scatter plot gave the energy of the particles, and this energy was used to determine the hadronic signal linearity of the calorimeter (Figure \ref{rd52}a). The angle $\theta$ (and thus the parameter $\chi$) were the same for all energies and types of hadron showers.

In practical calorimeters, the angle $\theta$ is located somewhere between 45$^\circ$ and 90$^\circ$. The larger the angle, \ie the larger the difference between $(h/e)_S$ and $(h/e)_C$, the better the dual-readout method works. This was demonstrated in detail by Groom \cite{deg13}, who pointed out the importance of this ``$h/e$ {\sl contrast}'' for dual-readout calorimeters. If $\theta = 45^\circ$, no complementary information is provided by the two types of signals. This is the situation one would encounter if the calorimeter was equipped with only one type of fibers (\eg scintillating ones), and these fibers were split into two bunches that are read out separately. In that case, the data points would cluster around the diagonal.
The fluctuations of the data points around that line would only be determined by event-to-event sampling differences between the two fiber bunches, and a rotation of the data points into the vertical plane would probably lead to a very narrow signal distribution. However, it would in that case not be possible to determine the coordinates of the pivot point, \ie the particle energy. The complementary information provided by the \v{C}erenkov signals makes it possible to determine these coordinates, and thus the particle energy, unambiguously.

The energy resolution of the RD52 dual-readout fiber calorimeter was thus determined as the width of the signal distribution for a collection of events produced by particles of the same energy, \ie a beam provided by a particle accelerator. No information about the composition or the energy of this beam was used. This is not different from the way the energy resolution is determined for other types of calorimeters, and therefore we use these resolutions for comparison (Figure \ref{Erescomp}).

\newpage

\bibliographystyle{unsrt}

\begin{thebibliography}{99.}

\bibitem{bgo} J.A.~Bakken \etal, \NIM {\bf A254} (1987) 535.

\bibitem{na48} G.D.~Barr \etal, \NIM {\bf A370} (1996) 413.

\bibitem{Liv17} Livan, M. and Wigmans, R. (2017). {\em Misconceptions about Calorimetry}, {\em Instruments} {\bf 2017} 1,3.
arXiv:1704.00661 [physics.ins-det].

\bibitem{Aco92b} D.~Acosta \etal, \NIM {\bf A316} (1992) 184.

\bibitem{Akc97} N.~Akchurin \etal, \NIM {\bf A399} (1997) 202.

\bibitem{Gab94} Gabriel, T.A. \etal,\NIM {\bf A338} (1994) 336.

\bibitem{Akc98}  N.~Akchurin \etal, \NIM {\bf A408} (1998) 380.

\bibitem{CMS07} N.~Akchurin \etal, {\em The response of CMS 
combined calorimeters to single hadrons, electrons and muons}, CERN-CMS-NOTE-2007-012.

\bibitem{Aba10} E. ~Abat \etal, \NIM {\bf A621} (2010) 134.

\bibitem{Bern87} E.~Bernardi \etal, \NIM {\bf A262} (1987) 229.

\bibitem{Aco91c} D.~Acosta \etal, \NIM {\bf A308} (1991) 481.

\bibitem{Suz99} T.~Suzuki \etal, \NIM {\bf A432} (1999) 48.

\bibitem{Beh90} U.~Behrens \etal, \NIM {\bf A289} (1990) 115. 

\bibitem{zeusjet} A.~Andresen \etal, \NIM {\bf A290} (1990) 95.

\bibitem{Abr81} H.~Abramowicz \etal, \NIM {\bf 180} (1981) 429.

\bibitem{Akc05a} N.~Akchurin \etal, \NIM  {\bf A537} (2005) 537.

\bibitem{deg07} D.E. Groom, \NIM {\bf A572} (2007) 633.

\bibitem{slee17} S.~Lee \etal, \NIM {\bf A866} (2017) 76.

\bibitem {Akc14} N.~Akchurin \etal, \NIM {\bf A735} (2014) 130.

\bibitem{Car16} A.~Cardini \etal, \NIM {\bf A808} (2016) 41.

\bibitem{geant} S. Agostinelli \etal, \NIM  {\bf A506} (2003) 250. 

\bibitem{Fritiof} B.~Andersson \etal ,  Nucl. Phys. {\bf B281} (1987) 289.

\bibitem{Bertini} D. Wright \etal, AIP Conf. Proc. {\bf 896} (2006) 11.

\bibitem{g4_pl}  http://geant4.cern.ch/support/proc{\_}mod{\_}catalog/physics{\_}lists/useCases.shtml

\bibitem{Akc09c} N.~Akchurin \etal, \NIM {\bf A610} (2009) 488.

\bibitem{Akc09a} N.~Akchurin \etal, \NIM {\bf A598} (2009) 422.





\bibitem{PDG16} C.~Patrignani \etal ~(Particle Data Group),  {\em Chin. Phys.} {\bf C40}, 100001 (2016), Section 34.9.2.





\bibitem{Dre90} G.~Drews \etal, \NIM {\bf A290} (1990) 335.

\bibitem{Aco91a} D.~Acosta \etal, \NIM {\bf A308} (1991) 481.

\bibitem{deg13} D.E.~Groom, \NIM {\bf A705} (2013) 24.























\end{thebibliography}

\end{document}